# L2T-Hyena: Enhancing State-Space Models with an Adaptive Learning to Teach Framework

F. Sohbati, Member, IEEE, F. Haddadi, and H. Salahinejad

***Abstract*— State-space models (SSMs) have recently emerged as efficient alternatives to computationally intensive architectures such as Transformers for sequence modeling. However, their training typically relies on static loss functions, which may be suboptimal at different stages of learning. In this work, we introduce a hybrid model that integrates the Hyena architecture with a Dynamic Loss Network (DLN) under a Learning-to-Teach (L2T) paradigm, referred to as L2T-DLN. In this framework, the Hyena model serves as a student whose loss function is adapted online, while a teacher model, equipped with a memory of the student's past performance, guides the DLN to dynamically trade off the primary cross-entropy objective and a regularization term. We evaluate the proposed L2T-Hyena model on the Penn Treebank (PTB) and WikiText-103 language modeling benchmarks and compare it against both a vanilla Hyena SSM and a Transformer baseline. On PTB, our model achieves a validation perplexity of 102.6, representing a substantial improvement over the 110.5 obtained by the vanilla Hyena trained with a static loss function and 121.28 achieved by the Transformer baseline. Similar gains are observed on WikiText-103, where L2T-Hyena reaches a validation perplexity of 68.3, outperforming vanilla Hyena (73.7) and Transformer (89.8). These results indicate that coupling SSMs with adaptive loss functions can significantly enhance both the quality and efficiency of deep learning models for sequential data and hold strong promise for applications in natural language processing, time-series analysis, and biological signal processing.**

## I. INTRODUCTION

Deep learning models, particularly Transformers, have achieved remarkable success in natural language processing tasks [1, 2, 3, 4, 5]. However, their architectures face a significant bottleneck when processing long sequences due to the self-attention mechanism's quadratic computational complexity $O(L^2)$ [6, 7, 8, 9]. The Hyena model, a State-Space Model (SSM), was introduced as a powerful alternative, leveraging long convolutional operations to reduce this complexity to a more manageable ($O(L\ log^L\ )$) while maintaining strong modeling capacity for long-range dependencies [10, 11].

Despite these architectural advantages, a major challenge in the optimization process persists due to the reliance on static loss functions. Such fixed loss functions cannot adapt to the evolving states of learning, which limits the model's ability to achieve optimal and rapid convergence. To address this issue, the Learning to Teach with a Dynamic Loss Network (L2T-DLN) framework provides an innovative solution by adjusting the loss function based on the model's performance during training, thereby enabling more intelligent and adaptive optimization [12, 13].

This research introduces the L2T-Hyena model, which combines the efficient Hyena architecture with the adaptive L2T-DLN framework. The primary objective is to evaluate this hybrid approach on word-level language modeling tasks using the Penn Treebank (PTB) and WikiText-103 datasets and to compare its effectiveness against two baselines: a vanilla Hyena model trained with a conventional static loss function and a Transformer-based language model of comparable capacity [10, 11, 12]. The main contributions of this paper are:

• Novel integration of the Hyena model with the L2T-DLN framework for adaptive loss optimization in a language modeling context.

• Comprehensive empirical evaluation on the PTB and WikiText-103 benchmarks, demonstrating that the proposed L2T-Hyena consistently achieves lower validation perplexity than both a vanilla Hyena SSM and a Transformer baseline.

• Detailed analysis of the dynamic interplay between the student model (Hyena), the teacher model, and the DLN, providing insights into how adaptive loss shaping improves training dynamics and generalization in sequence modeling.

## II. RELATED WORK

This section provides an overview of the core concepts that form the foundation of our proposed model: State-Space Models (SSMs), the Hyena architecture, and Learning to

(Corresponding author: Farzan Haddadi.)
F. Sohbati is with the School of Electrical Engineering, Iran University of Science and Technology, Tehran, Iran (e-mail: fatemehsohbati2710@gmail.com).
F. Haddadi is with the School of Electrical Engineering, Iran University of Science and Technology, Tehran, Iran (e-mail: farzanhaddadi@iust.ac.ir).
H. Salahinejad is with the School of Computer Engineering, Tarbiat Modares University, Tehran, Iran (e-mail: hamidsalahinejad25@gmail.com).

Teach with Dynamic Loss Networks (L2T-DLN) framework [9,10,12].

*A. State Space Models*

State-Space Models (SSMs) are a class of mathematical models that describe a dynamic system using a set of first-order differential or difference equations. The core of an SSM is the state vector, a multi-dimensional vector that summarizes the past history of the system [14,15].

For a linear continuous-time system, the SSM is defined by two primary equations:

1. State Equation, which describes how the state vector x(t) evolves over time based on current state and an input vector u(t):

$$\dot{x}(t) = Ax(t) + Bu(t) \qquad (1)$$

Where, $A$ is the state evolution matrix that governs the system's internal dynamics, and $B$ is the input transformation matrix that determines the influence of the input on the state [14,15].

2. The Output Equation, which defines how the output vector y(t) is generated from the current state and input:

$$y(t) = Cx(t) + Du(t) \qquad (2)$$

Where, $C$ is the output matrix that maps the state to the output, and $D$ is the feedthrough matrix that models the direct impact of the input on the output [14,15].

In deep learning, SSMs have become a promising alternative to attention-based architectures, as their structure is highly effective for processing time-series data. A key advantage is their linear computational complexity with respect to sequence length, a significant improvement over the quadratic complexity of attention mechanisms, making them exceptionally efficient for long sequences. Notable SSM architectures include S4 and Mamba [9,15,16,17].

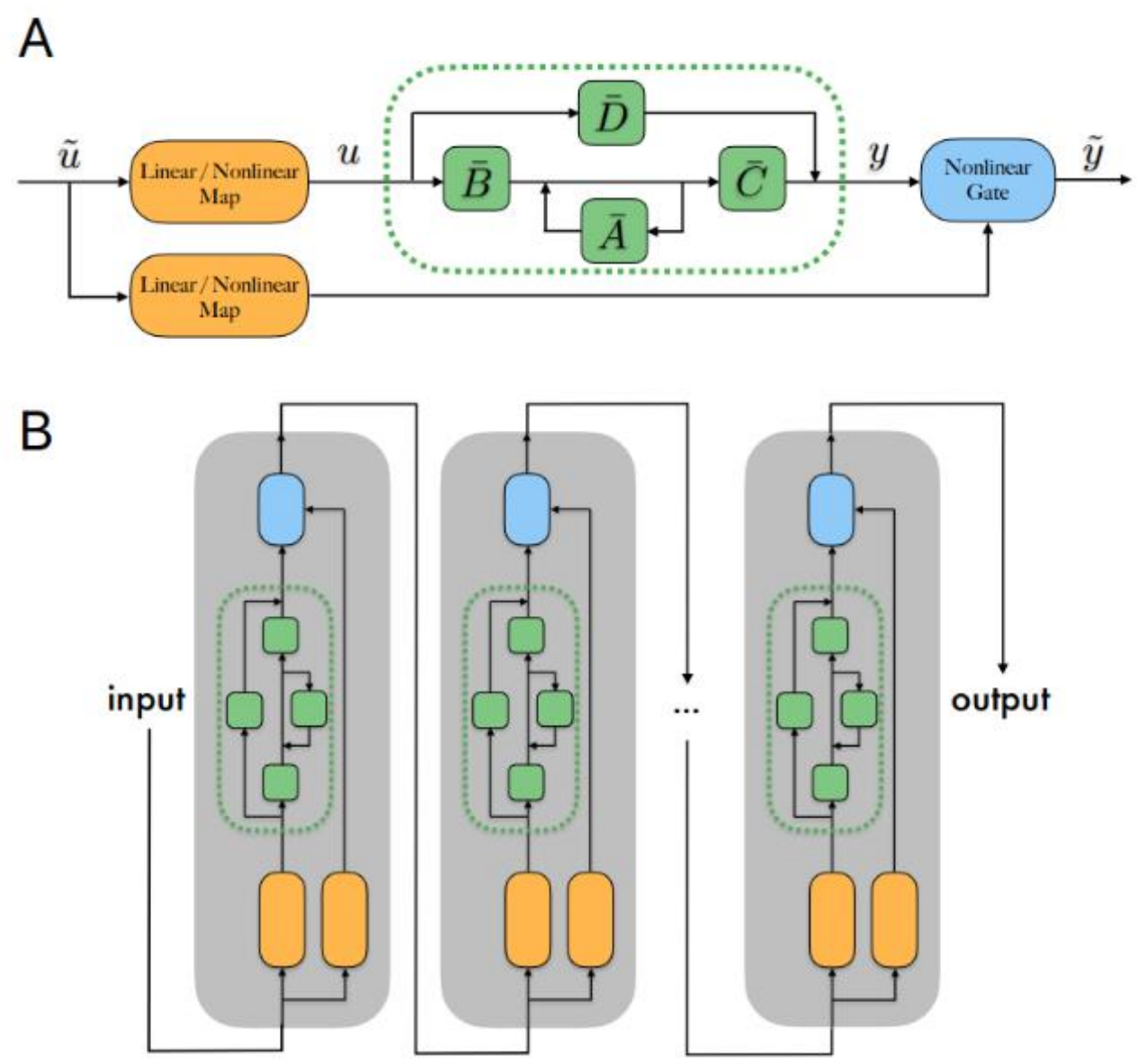

**Fig. 1.** A. General scaffolding of an SSM. The dynamical model is represented in green. The input to the SSM is pre-processed and forked off in a skip connection (lower signal). The nature of the pre-processing map (linear or nonlinear) depends on the specific scaffolding. The output of the recursion is then post-processed with a nonlinear gate. B. Overall architecture of an SSM. Each of the SSMs including its scaffolding is included in a layered fashion, where the output from one layer is the input to the next [14].

*B. The Hyena Model*

The Hyena model was proposed as a scalable, low-cost alternative to the attention mechanism in Transformers, specifically addressing its $O(L^2)$ computational complexity. Hyena models long-range dependencies with sub-quadratic complexity $(O(L\ log^{L}))$ by employing a combination of implicit long convolutions and data-dependent multiplicative gating. Instead of an attention layer, Hyena parameterizes its convolutional filters implicitly using a small feed-forward network and leverages Fast Fourier Transforms (FFTs) for highly efficient computation [10,11,18,19].

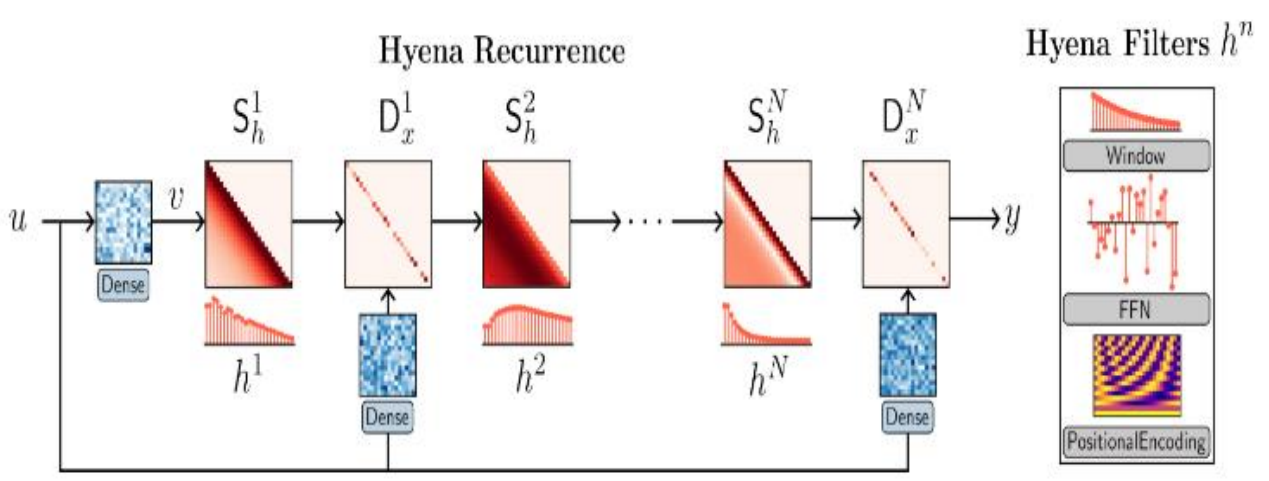

**Fig. 2.** Hyena operator is defined as a recurrence of two efficient subquadratic primitives: an implicit long convolution h (i.e. Hyena filters parameterized by a feed-forward network) and multiplicative elementwise gating of the input. The depth of the recurrence specifies the size of the operator. Hyena can equivalently be expressed as a multiplication with data-controlled (conditioned by the input u) diagonal matrices $D_x$ and Toeplitz matrices $S_h$ In addition, Hyena exhibits sublinear

parameter scaling (in sequence length) and unrestricted context, similar to attention, while having lower time complexity [10].

### C. Learning to Teach with Dynamic Loss Networks

The Learning to Teach (L2T) paradigm is a meta-learning approach that aims to optimize a student model's training process through the guidance of a teacher model. L2T-DLN is an advanced adaptive optimization framework that uses a dynamic network to guide the student model toward better performance across different learning stages.

L2T-DLN features a memory-augmented teacher model, which allows the system to learn from accumulated experiences during training. The framework involves a strong, dynamic interaction between three components [12,13]:

- The student model is trained using the loss weights provided by the DLN.
- The Dynamic Loss Network (DLN) is updated based on feedback from the teacher.
- The teacher model, equipped with memory units, learns to provide optimal guidance by retaining information about long-term training dependencies

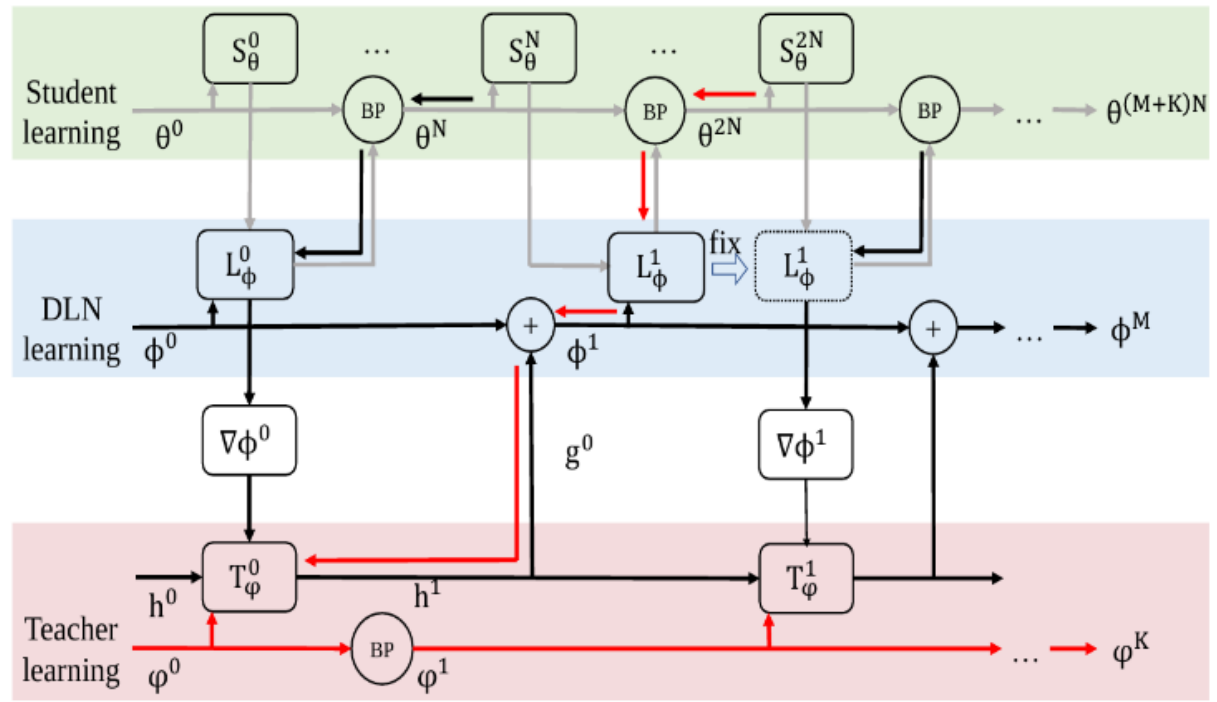

**Fig. 3.** The pipeline of L2T-DLN [12]

## III. Methodology

The proposed model, L2T-Hyena, integrates the Hyena architecture as a student model within a Learning to Teach (L2T) framework. This framework employs a Dynamic Loss Network (DLN) for adaptive loss optimization, aiming to enhance language modeling by combining Hyena's efficient sequence modeling capabilities with an adaptive learning mechanism [10,12,13]. In the following, we first describe the student model based on Hyena, then present the teacher and DLN components of the L2T framework, and finally provide an architectural overview of the complete system.

### A. Student Model (Hyena)

The student model is a Hyena-based architecture designed for word-level language modeling. It begins with an embedding layer that maps input tokens from a 10,000-token vocabulary to 256-dimensional vectors, which are then combined with positional embeddings. The core of the model is a stack of 6 Hyena blocks that capture long-range dependencies using a combination of a short convolution (kernel size 3) and a long, FFT-based convolution whose filter is generated dynamically. Key architectural parameters include a model dimension of 256 and a Hyena filter order of 2. Finally, an output projection layer maps the processed sequence back to the vocabulary size to produce logits for next-token prediction. The weights of this final layer are tied to the input embedding matrix in order to improve parameter efficiency and regularization [2, 10, 20].

### B. Dynamic Loss Network

The primary role of the Dynamic Loss Network (DLN) is to adaptively adjust the weighting between the standard cross-entropy loss and an L2 logit regularization term. For each training instance, it takes the student's logits and target tokens as input and extracts a set of five statistical features that capture aspects such as prediction confidence and error magnitude. This feature sequence is first normalized and then fed into a Gated Recurrent Unit (GRU), which summarizes it into a single representative hidden state vector. The summary vector is subsequently passed through a 4-layer multilayer perceptron (MLP) that produces a raw scalar weight for the regularization term. Finally, a sigmoid activation is applied to this scalar to obtain the final weight in the range (0, 1), which is used to balance the cross-entropy and regularization components of the overall loss [12, 21, 22, 23].

### C. Teacher Model and L2T Framework

The teacher model is a memory-augmented system whose core predictive component is a multilayer perceptron (MLP) that guides the learning process of the DLN. Instead of processing text directly, it operates on a memory buffer that stores up to 500 past experiences. Each experience contains the statistical features generated by the DLN for a given sequence, along with the corresponding student loss. During its training step, the teacher strategically samples from this memory, prioritizing high-loss examples in order to focus on more challenging cases.

Given a proposed regularization weight from the DLN, the teacher's task is to predict the resulting student loss. It is optimized by minimizing the Huber loss between its prediction and the actual student loss, which makes the training robust to outliers and stabilizes the learning dynamics of the overall L2T framework [24].

### D. Architectural Overview

Figure 4 illustrates the complete training pipeline of the proposed L2T-Hyena framework. The process begins with data preprocessing (tokenization and token-to-ID conversion). The student model (Hyena) processes the input and produces logits $z_s$. The Dynamic Loss Network (DLN) takes these logits along with the true labels $y$ and outputs a scalar weight $w_{reg} \in [0.05,0.95]$ for the regularization term. The total loss is computed as a weighted combination of cross-entropy and L2 regularization, with gradients used to update the student model. Simultaneously, features extracted by the DLN and the corresponding true cross-entropy loss are stored in the teacher's memory. The teacher model samples from this memory (prioritizing high-loss examples) and, given a proposed $w_{reg}$, predicts the expected student loss. The teacher is trained using a Huber loss objective, and its feedback is used to meta-update the DLN. This closed-loop interaction enables the system to dynamically adapt the loss function based on the student's ongoing performance.

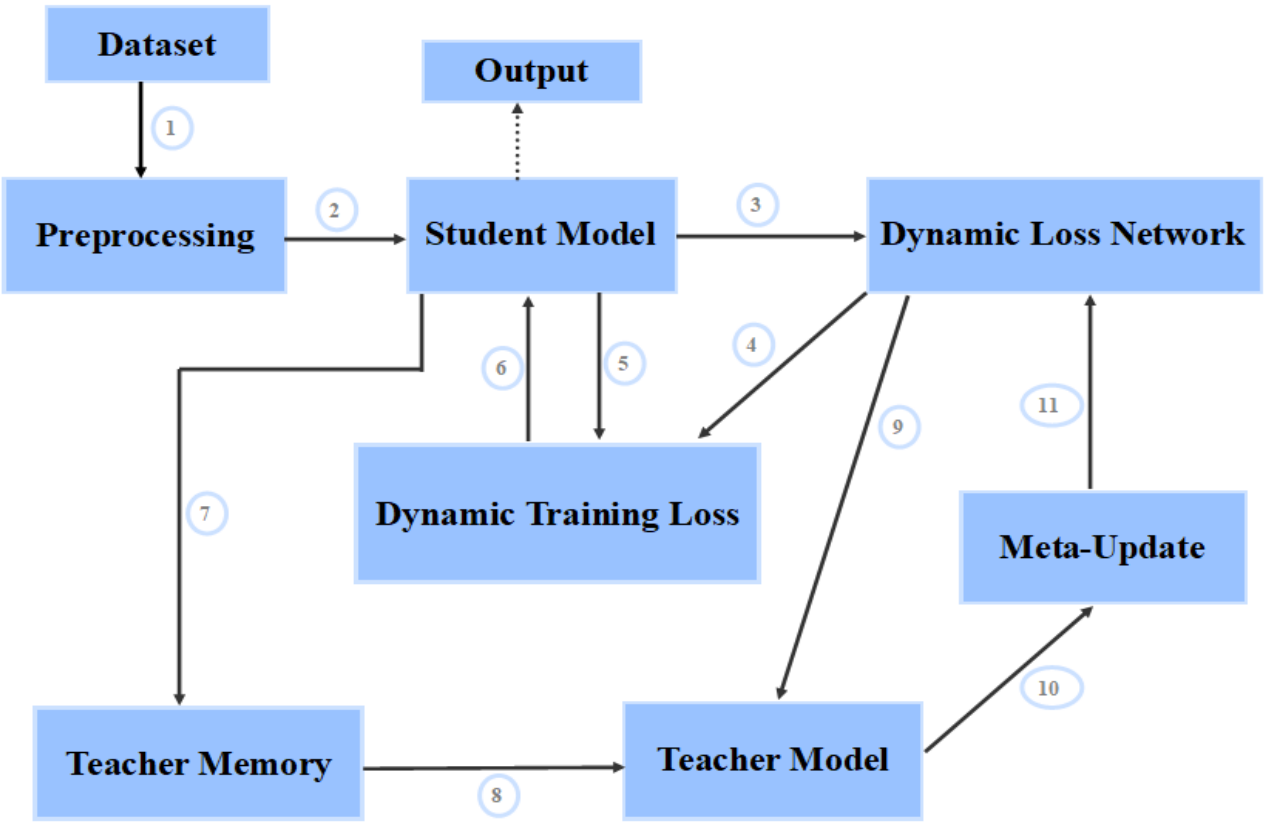

**Fig. 4.** Overview of the proposed L2T-Hyena framework integrating the Hyena student model, Dynamic Loss Network (DLN), and memory-augmented teacher model.

## IV. EXPERIMENTAL SETUP

This section details the datasets, implementation specifics, and the baseline models used for comparative analysis.

### *A. Dataset*

We evaluate our models on two widely used word-level language modeling benchmarks: Penn Treebank (PTB) and WikiText-103.

Penn Treebank (PTB) is a standard benchmark for language modeling, consisting of approximately 930,000 training tokens and 74,000 validation tokens, with a vocabulary of 10,000 unique tokens. We follow the conventional preprocessing and data splits commonly used in prior work.

WikiText-103 is a large-scale language modeling corpus constructed from high-quality Wikipedia articles. It contains long, coherent documents with richer long-range dependencies compared to PTB, making it well suited for evaluating sequence models on more realistic text. We use the standard train/validation/test splits and apply the same tokenization pipeline as in PTB, while allowing for a larger effective context window to better exploit the longer sequences.

### *B. Implementation Details*

All models are implemented in PyTorch using a standard mini-batch training loop. Each of the three primary components the student, teacher, and DLN is optimized using a separate AdamW optimizer with distinct hyperparameters. Unless otherwise stated, these settings are used for experiments on both the PTB and WikiText-103 datasets.

The loss functions are tailored to each component's role. The student model is trained with a dynamically weighted combination of cross-entropy and an L2 logit regularization term, where the weight is produced by the DLN. Both the teacher and the DLN are trained with their respective objective functions described in Section III.

For all experiments, models are trained for 10 epochs with a batch size of 128 and a sequence length of 64. A cosine-annealing learning rate scheduler with a 2-epoch warm-up period is used to ensure stable convergence. For additional stability, gradient clipping is applied to all models during training. The same training pipeline (optimizer type, batch size, sequence length, and learning-rate schedule) is used for both the baselines and the proposed L2T-Hyena model to ensure a fair comparison.

### *C. Baseline Models*

For a fair and direct comparison, we consider two baseline models:

1) Vanilla Hyena: A standalone Hyena model trained without the L2T-DLN framework. It uses the same architecture as the student model (embedding dimension, number of Hyena blocks, and vocabulary size) and is optimized with the same hyperparameters as the student model. The training objective is a standard cross-entropy loss with a fixed L2 logit regularization term, without any adaptive weighting.

2) Transformer: A Transformer-based language model serving as a widely used baseline for sequence modeling. It is configured to have a comparable number of trainable parameters and computational cost to the Hyena-based student model, and it is trained using the same data preprocessing, batch size, sequence length, optimizer, and learning-rate schedule as the Hyena models. This setup ensures that performance differences primarily reflect architectural and training-strategy choices rather than discrepancies in optimization.

## V. RESULTS AND DISCUSSION

This section presents the experimental results of the proposed L2T-Hyena model and compares them with a vanilla Hyena model and a Transformer baseline. The analysis is performed on both the PTB and WikiText-103 datasets and includes quantitative metrics, qualitative evaluation of generated text, and a visual study of the training dynamics.

### *A. Results on The PTB Dataset*

Table1 reports the PTB results for the three models under consideration: the proposed L2T-Hyena model, a vanilla Hyena SSM, and a vanilla Transformer baseline. For each model, the table lists the epoch at which the best validation performance is achieved, together with the corresponding training and validation loss, as well as training and validation perplexity.

TABLE I

TRAINING RESULTS OF THREE MODELS ON THE PTB DATASET

| Models | Epoch | Perplexity | Loss |
|---|---|---|---|
| L2T-Hyena model | 5 | 102.60 | 4.63 |
| vanilla Hyena | 3 | 110.45 | 4.70 |
| Transformer | 2 | 121.28 | 4.80 |

As shown in Table1, the proposed L2T-Hyena model attains the lowest validation perplexity (102) at epoch 5, outperforming both the vanilla Hyena (validation perplexity 110 at epoch 3) and the Transformer baseline (validation perplexity 121 at epoch 2). This improvement suggests that the dynamic loss adjustment provided by the L2T framework leads to more effective optimization and better generalization on the PTB dataset.

In addition, the gap between training and validation metrics is less pronounced for L2T-Hyena than for the baselines, particularly when compared to the Transformer model, which exhibits the largest discrepancy between training and validation perplexity. This pattern is indicative of a stronger tendency to overfit in the baseline models, whereas the proposed L2T-Hyena model maintains a more favorable trade-off between fitting the training data and preserving validation performance.

In addition to the quantitative results, we also perform a qualitative comparison of text generated by the three models on the PTB dataset. For a shared prompt, the vanilla Hyena model generally produces sentences that are locally grammatical but sometimes lack global coherence. The Transformer baseline tends to generate more complex sentence structures but may exhibit repetitive patterns or semantic drift. In contrast, the proposed L2T-Hyena model produces continuations that are typically more coherent and semantically consistent, which is in line with the quantitative improvements reported above.

TABLE II

QUALITATIVE COMPARISON OF TEXT GENERATED BY THE THREE MODELS ON THE PTB DATASET

| Models | Best Perplexity | The Generated Text |
|---|---|---|
| L2T-Hyena model | 102.60 | the meaning of life is entirely good says washington <unk> professor of killed at simpson <unk> |
| vanilla Hyena | 110.45 | the meaning of life is the party who has to encourage the idea for months he belongs to the outspoken having right to make it |
| Transformer | 121.28 | the meaning of life is about Pollock with the jury was admit from the evidence of Wood 's historical life. |

Figure 5 depicts the validation perplexity over training epochs for the proposed L2T-Hyena model, the vanilla Hyena model, and the Transformer model on the PTB dataset. The curves in Figure 5 clearly show that the L2T-Hyena model achieves the lowest validation perplexity and maintains a more stable trajectory, whereas the baselines reach their optimal points earlier and then deteriorate, confirming the effectiveness of the adaptive loss strategy in delaying and reducing overfitting.

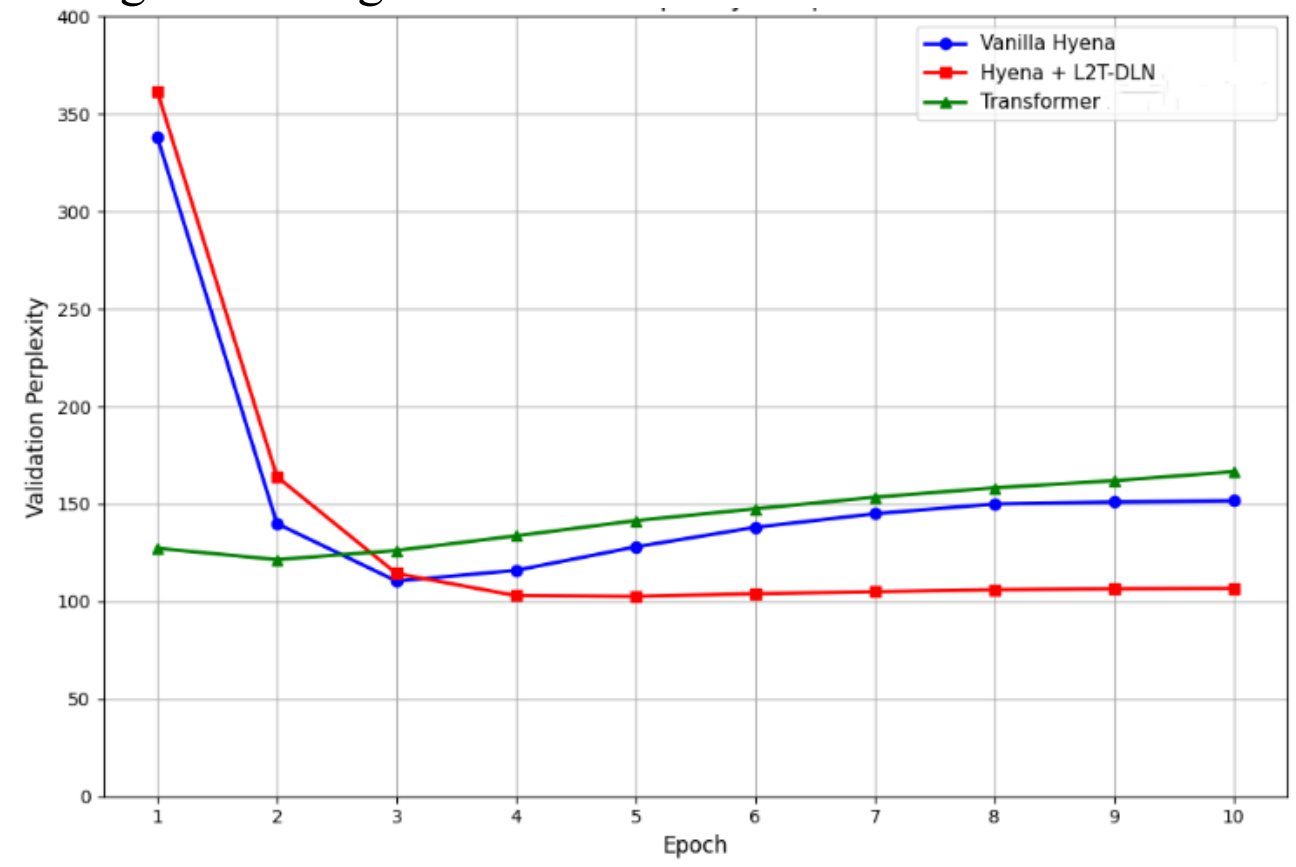

**Fig. 5.** Comparison of validation perplexity trajectories during training for the three models on the PTB dataset

Figure 6 illustrates the training cost for the three models on PTB. All models exhibit decreasing training cost over time, but the curve corresponding to the proposed model converges more smoothly and to a lower final value. When Figures 5 and 6 are considered together, they indicate that the L2T-Hyena model not only optimizes the training objective efficiently but also translates this improvement into consistently better validation performance compared to the vanilla Hyena and Transformer baselines.

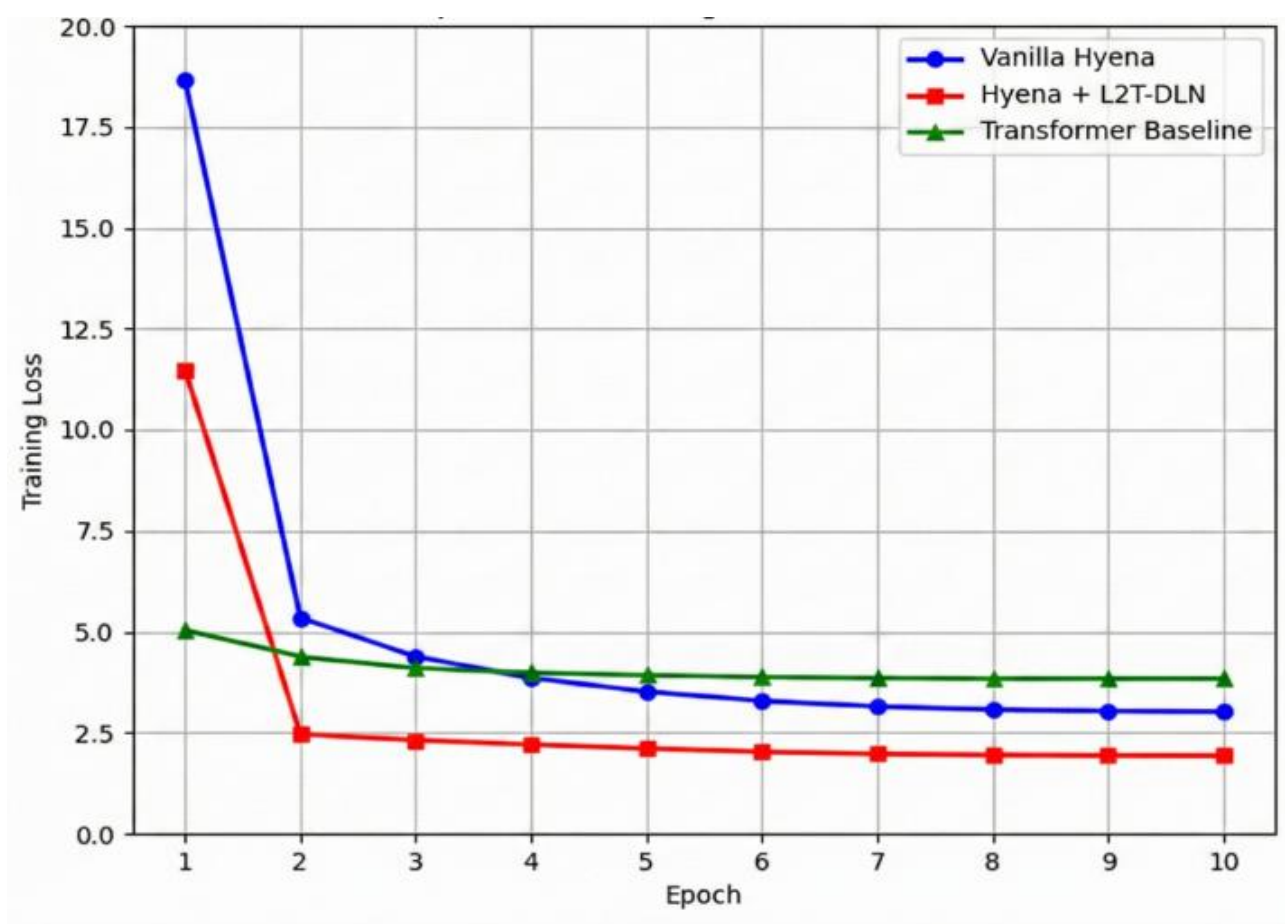

**Fig. 6.** Comparison of training loss on the PTB dataset

*B. Results on The Wiki Text-103 Dataset*

Table3 summarizes the WikiText-103 results for the three models: the proposed L2T-Hyena model, the vanilla Hyena model, and a vanilla Transformer baseline. For each model, the table reports the loss and validation perplexity at epoch 30.

TABLE III

TRAINING RESULTS OF THREE MODELS ON THE WIKI TEXT-103 DATASET

| Models | Epoch | Perplexity | Loss |
|---|---|---|---|
| L2T-Hyena model | 30 | 65.5 | 4.13 |
| vanilla Hyena | 30 | 72.1 | 4.23 |
| Transformer | 30 | 85.0 | 4.26 |

As shown in Table 3, the proposed L2T-Hyena model achieves the lowest validation loss (4.13) and the lowest perplexity (65.5) after 30 epochs, outperforming both the vanilla Hyena model (loss 4.23, perplexity 72.1) and the Transformer baseline (loss 4.26, perplexity 85.0). These results confirm that the benefits of the L2T-based dynamic loss adjustment extend to larger and more complex corpora such as WikiText-103, and that the L2T-Hyena model maintains a clear advantage over both static-loss Hyena and Transformer architectures throughout extended training.

TABLE IV

QUALITATIVE COMPARISON OF TEXT GENERATED BY THE THREE MODELS ON THE WIKI TEXT-103DATASET

| Models | Best Perplexity | The Generated Text |
|---|---|---|
| L2T-Hyena model | 65.5 | the meaning of life is one in Scotland and elsewhere. The independent dates of papers follow a ladder, as from a number of thrones. |
| vanilla Hyena | 72.1 | the meaning of life is about Pollock, where the jury was admitted from the evidence of Wood's historical life. It was <unk> carried. |
| Transformer | 85.0 | the meaning of life is the danger of misery that causes more pain, which in another case turns into a dialog. |

Table 4 reports a qualitative comparison of text generated by the three models on the WikiText-103 dataset, along with their corresponding best validation perplexities. For a shared prompt ("the meaning of life is …"), the Transformer baseline produces a continuation that, while grammatically structured, introduces concepts ("danger of misery", "pain", "dialog") that are semantically disjointed and lack clear coherence with the prompt. The vanilla Hyena model generates a sentence that begins on-topic but quickly drifts into a specific named-entity context ("Pollock", "Wood's historical life") and includes an incomplete token (<unk>), reducing readability. In contrast, the proposed L2T-Hyena model yields a continuation that stays closer to the prompt's theme, maintains grammatical coherence, and avoids abrupt topical shifts or artifacts, indicating that the improvements in perplexity are accompanied by perceptibly higher text quality and better prompt adherence.

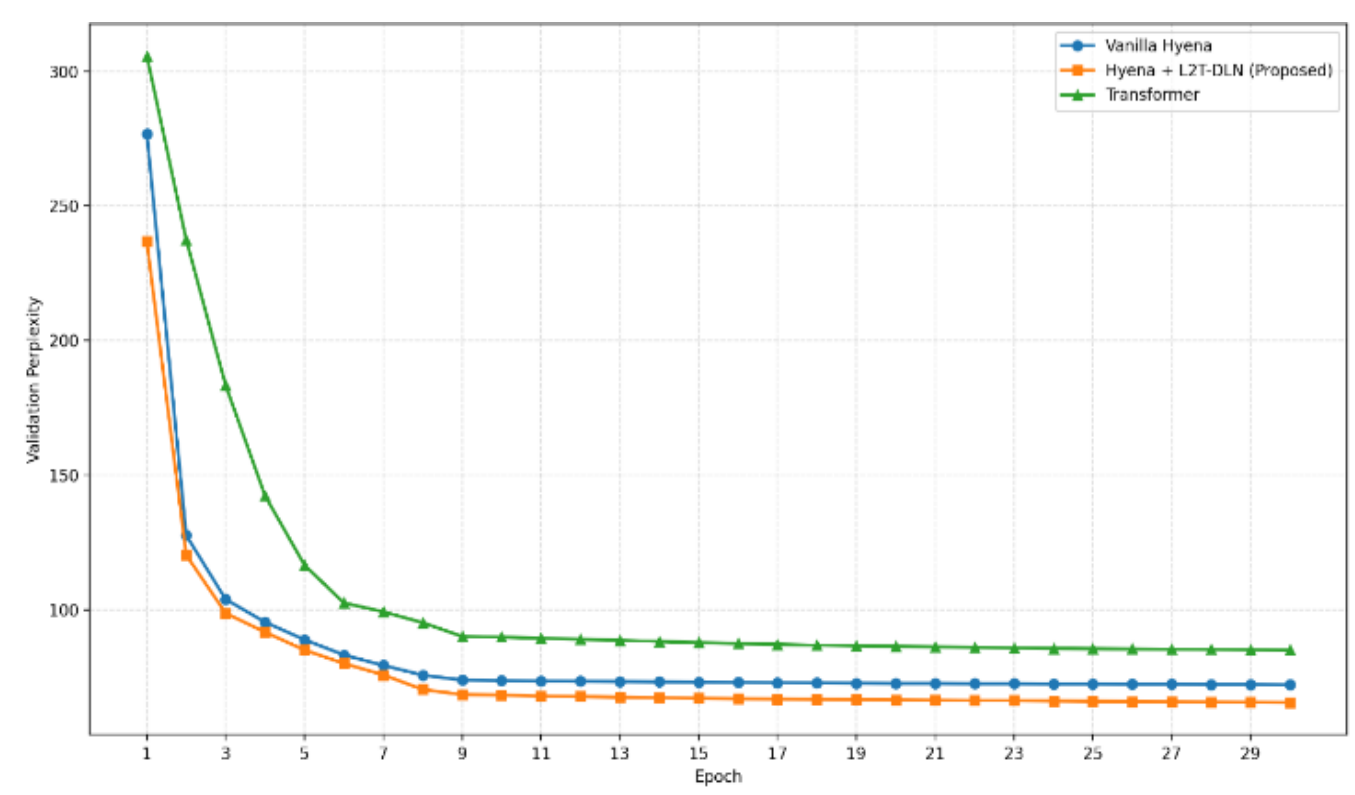

**Fig. 7.** Comparison of validation perplexity trajectories during training for the three models on the Wiki Text-103 dataset.

Figure 7 illustrates the evolution of validation perplexity over 30 training epochs on the WikiText-103 dataset for the proposed L2T-Hyena model, the vanilla Hyena model, and the Transformer baseline. Across almost all epochs, L2T-Hyena attains the lowest validation perplexity, with a consistent margin over the vanilla Hyena model. Both Hyena-based models substantially outperform the Transformer throughout the entire training process. All three models exhibit steadily decreasing perplexity that gradually plateaus toward later epochs, indicating stable convergence on this large-scale corpus. The proposed model not only converges to a lower final perplexity but also shows a smoother and more stable descent, underscoring the effectiveness of the adaptive loss-weighting mechanism in enhancing generalization.

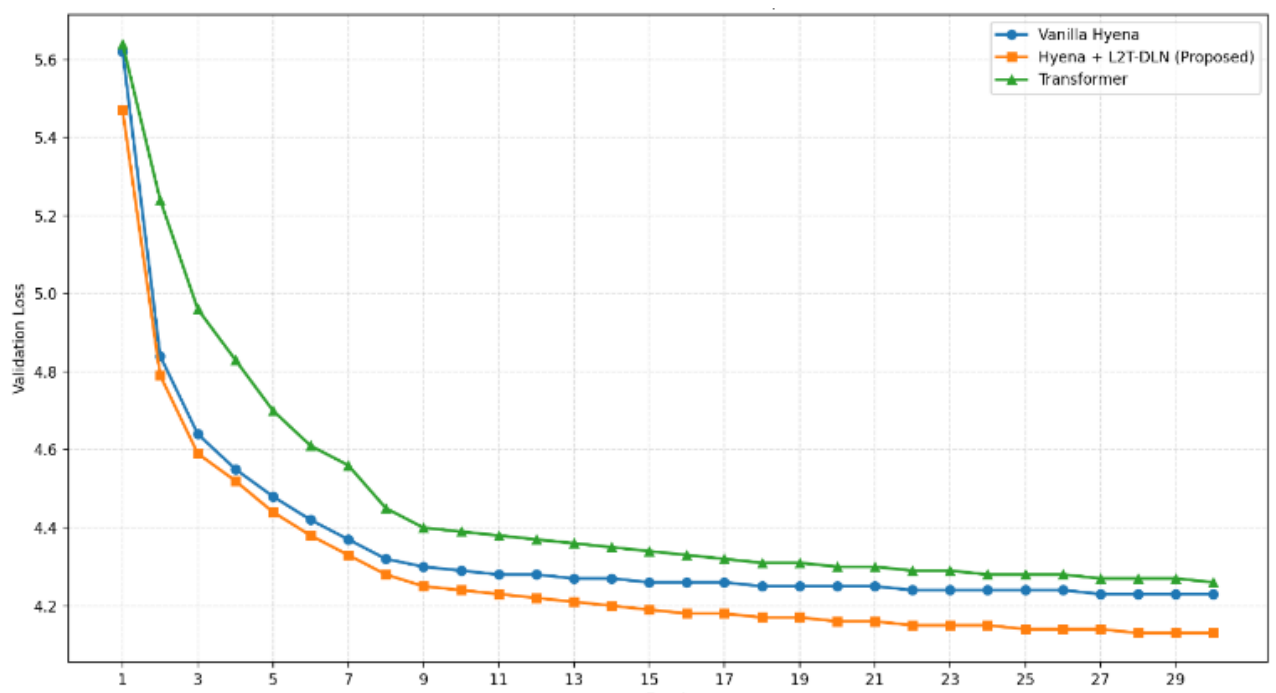

**Fig. 8.** Comparison of training loss on the Wiki Text-103 dataset

Figure 8 shows the training loss curves over 30 epochs for the three models on the WikiText-103 dataset. The proposed L2T-Hyena model maintains a consistently lower training loss across nearly all epochs, converging smoothly to a final loss of approximately 3.91 by epoch 30. The vanilla Hyena model also shows a descending loss trend but remains slightly higher (≈ 4.02), while the Transformer baseline plateaus at a higher loss (≈ 4.31). The smoother and lower loss trajectory of L2T-Hyena indicates more stable optimization and reduced overfitting compared to both baselines. Taken together with the validation perplexity trends in Figure 7, these results demonstrate that the adaptive loss-weighting mechanism not only lowers training loss more effectively but also translates this advantage into substantially better generalization performance.

## VI. Conclusion and Future Work

This research has demonstrated that integrating the Hyena architecture with the L2T-DLN framework can significantly enhance language-modeling performance. Evaluated on both the PTB and WikiText-103 datasets, the proposed L2T-Hyena model consistently outperforms a standalone Hyena model trained with a static loss function and a Transformer baseline of comparable capacity. On PTB, L2T-Hyena reduces the validation perplexity from 110.5 (vanilla Hyena) and 144.1 (Transformer) to 102.6, while on WikiText-103 it improves perplexity from 72.1 and 85.0 to 65.5 after 30 training epochs. These results highlight the efficacy of combining state-space architectures with adaptive loss optimization and guided learning strategies to better unlock the potential of efficient sequence models.

The primary novelty of this work lies in the integration of the Hyena architecture with the L2T-DLN framework in a language-modeling setting, establishing an effective training paradigm that jointly exploits the strengths of state-space models and learning-to-teach techniques.

Future research could explore several directions:

- **Broader evaluation:** Extending L2T-Hyena to larger and more diverse language-modeling benchmarks, as well as to downstream NLP tasks such as machine translation and text summarization, in order to further assess its generalizability in practical applications.

- **Deeper framework analysis:** Conducting a more in-depth analysis of the L2T-DLN components, for example by investigating the impact of varying the teacher model's memory size and sampling strategies on the training dynamics, or by exploring more advanced architectures for the teacher (e.g., LSTM- or Transformer-based teachers) to better model the student's learning history.

- **Comparative studies:** Benchmarking L2T-Hyena against a wider range of state-of-the-art language models, including more advanced SSMs (e.g., Mamba), larger-scale Transformer variants, and hybrid architectures that combine attention and state-space mechanisms, to better position this work within the current landscape.

## References

[1] M. Shao, A. Basit, R. Karri and M. Shafique, "Survey of Different Large Language Model Architectures: Trends, Benchmarks, and Challenges," *in IEEE Access*, vol. 12, pp. 188664-188706, doi: 10.1109/ACCESS.2024.3482107, 2024.

[2] J. Devlin, M.-W. Chang, K. Lee, and K. Toutanova, "BERT: Pre-training of Deep Bidirectional Transformers for Language Understanding," *in Proc. NAACL-HLT*, 2019.

[3] T. B. Brown, B. Mann, N. Ryder, M. Subbiah, J. Kaplan, P. Dhariwal, and D. Amodei, "Language Models are Few-Shot Learners," *in Proc. Adv. Neural Inf. Process. Syst. (NeurIPS),* 2020.

[4] C. Raffel, N. Shazeer, A. Roberts, K. Lee, S. Narang, M. Matena, and P. J. Liu, "Exploring the Limits of Transfer Learning with a Unified Text-to-Text Transformer," *J. Mach. Learn. Res.*, vol. 21, pp. 1–67, 2020.

[5] S. Minaee, T. Mikolov, N. Nikzad, M. Asgari Chenaghlu, R. Socher, X. Amatriain, and J. Gao, "Large Language Models: A Survey," *arXiv preprint* arXiv:2402.06196, 2024.

[6] H. Naveed, A. U. Khan, S. Qiu, M. Saqib, S. Anwar, M. Usman, N. Barnes, and A. S. Mian, "A Comprehensive Overview of Large Language Models*," arXiv preprint* arXiv:2307.06435, 2023.

[7] S. B, P. R. P, S. M. B and K. S, "The Evolution of Large Language Model: Models, Applications and Challenges," *International Conference on Current Trends in Advanced Computing (ICCTAC)*, Bengaluru, India, pp. 1-8, doi: 10.1109/ICCTAC61556.2024.10581180, 2024.

[8] M. A. K. Raiaan, M. S. H. Mukta, K. Fatema, N. M. Fahad, S. Sakib, M. M. J. Mim, J. Ahmad, M. E. Ali, and S. Azam, "A Review on Large Language Models: Architectures, Applications, Taxonomies, Open Issues and Challenges," *IEEE Access*, vol. 12, pp. 26839–26874, 2024.

[9] X. Wang, S. Wang, Y. Ding, Y. Li, W. Wu, Y. Rong, W. Kong, J. Huang, S. Li, H. Yang, Z. Wang, B. Jiang, C. Li, Y. Wang, Y. Tian, and J. Tang, “State Space Model for New-Generation Network Alternative to Transformers: A Survey,” *arXiv preprint* arXiv:2404.09516, 2024.

[10] M. Poli, S. Massaroli, E. Nguyen, D. Y. Fu, T. Dao, S. Baccus, Y. Bengio, S. Ermon, and C. Ré, “Hyena Hierarchy: Towards Larger Convolutional Language Models,” *in Proceedings of the 40th International Conference on Machine Learning (ICML'23),* vol. 202, pp. 28043–28078, 2023.

[11] T. R. Ralambomihanta, S. Mohammadzadeh, M. S. N. Islam, W. Jabbour, and L. Liang, “Scavenging Hyena: Distilling Transformers into Long Convolution Models,” *arXiv preprint.* arXiv:2401.17574, 2024.

[12] Z. Hai, L. Pan, X. Liu, Z. Liu, and M. Yunita, "L2T-DLN: Learning to teach with dynamic loss network," *in Proceedings of the 37th International Conference on Neural Information Processing Systems (NeurIPS '23),* Red Hook, NY, USA, 2023.

[13] L. Wu, F. Tian, Y. Xia, Y. Fan, T. Qin, J. Lai, and T.-Y. Liu, “Learning to Teach with Dynamic Loss Functions,” *Neural Information Processing Systems*. arXiv:1810.12081, 2018.

[14] C. A. Alonso, J. Sieber, and M. N. Zeilinger, “State Space Models as Foundation Models: A Control Theoretic Overview,” *arXiv preprint* arXiv:2403.16899, 2024.

[15] A. Gu, K. Goel, and C. Ré, “Efficiently Modeling Long Sequences with Structured State Spaces,” *arXiv preprint* arXiv:2111.00396, 2021.

[16] P. Baldi, “Autoencoders, Unsupervised Learning, and Deep Architectures,” *JMLR: Workshop and Conference Proceedings*, vol. 27, pp. 37–50, 2012.

[17] B. N. Patro and V. S. Agneeswaran, "Mamba-360: Survey of State Space Models as Transformer Alternative for Long Sequence Modelling: Methods, Applications, and Challenges," *arXiv preprint* arXiv:2404.16112, 2024.

[18] T. Kudo, “Subword Regularization: Improving Neural Network Translation Models with Multiple Subword Candidates,” *in Proc. Assoc. Comput. Linguistics (ACL),* pp. 66–75, 2018.

[19] V. Nair and G. E. Hinton, “Rectified Linear Units improve Restricted Boltzmann Machines,” *in Proc. Int. Conf. Mach. Learn. (ICML)*, pp. 807–814,2010.

[20] R. Sennrich, B. Haddow, and A. Birch, “Neural Machine Translation of Rare Words with Subword Units,” *in Proc. Assoc. Comput. Linguistics (ACL),* pp. 1715–1725, 2016.

[21] I. Sutskever, O. Vinyals, and Q. V. Le, “Sequence to Sequence Learning with Neural Networks,” *in Proc. Adv. Neural Inf. Process. Syst. (NeurIPS),* 2014.

[22] D. Hendrycks and K. Gimpel, “Gaussian Error Linear Units (GELUs),” *arXiv preprint* arXiv:1606.08415, 2016.

[23] J. L. Ba, J. R. Kiros, and G. E. Hinton, “Layer Normalization,” arXiv preprint arXiv:1607.06450, 2016.

[24] L. Ouyang, J. Wu, X. Jiang, D. Almeida, C. L. Wainwright, P. Mishkin, C. Zhang, S. Agarwal, K. Slama, A. Ray, J. Schulman, J. Hilton, F. Kelton, L. Miller, M. Simens, A. Askell, P. Welinder, P. Christiano, J. Leike, and R. Lowe, “Training language models to follow instructions with human feedback,” *in Advances in Neural Information Processing Systems (NeurIPS),* 2022.

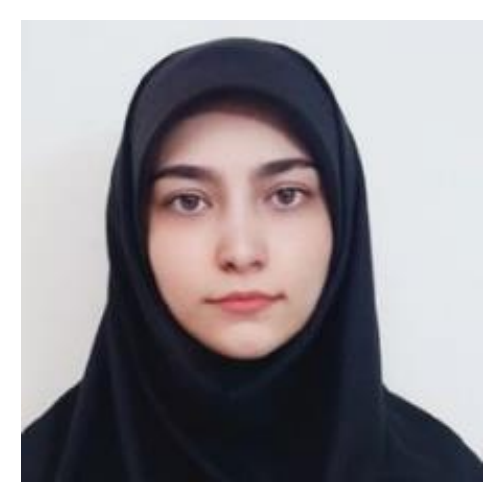

**FATEMEH SOHBATI** is currently pursuing the M.Sc. degree in communication systems from the Iran University of Science and Technology, Tehran, Iran. Her research interests include language models and deep learning.

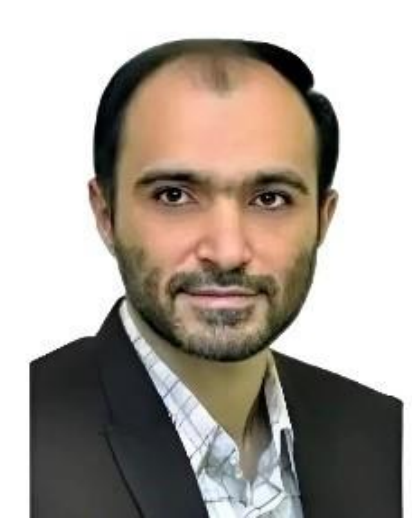

**FARZAN HADDADI** received the Ph.D. degree in communication systems from Sharif University of Technology, Tehran, Iran. He is currently a University Professor with the School of Electrical Engineering/Department of Communication Systems, Iran University of Science and Technology, Tehran, Iran. His research interests include deep learning, neural networks, and artificial intelligence.

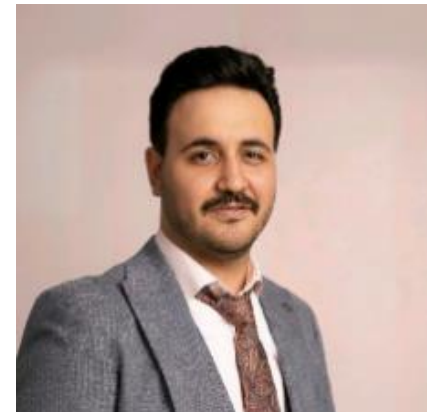

**HAMID SALAHINEJAD** is currently pursuing the M.Sc. degree in secure computing from Tarbiat Modares University, Tehran, Iran. His research interests include language models and the Ethereum network.